\documentclass[preprint,12pt]{elsarticle}

\usepackage{comment}
\usepackage{siunitx}
\usepackage{upgreek}
\usepackage{subcaption}
\usepackage[colorlinks=false,linkcolor=black, 
            citecolor=black,urlcolor=black]{hyperref}
\usepackage{lineno}
\usepackage{amsmath}

\usepackage{xcolor}

\journal{Nuclear Instruments and Methods in Physics Research A}

\begin{document}
\begin{frontmatter}

\title{First evaluation of the tracking performance of the ARCADIA Fully Depleted MAPS with high-energy proton beam}

\affiliation[dfa]{
    organization={University of Padova, Department of Physics and Astronomy "G. Galilei"},
    city={Padova},
    country={Italy}
}
\affiliation[cisas]{
    organization={University of Padova, Centro di ateneo di Studi e Attività Spaziali CISAS "G. Colombo"},
    city={Padova},
    country={Italy}
}
\affiliation[cassino]{
    organization={University of Cassino and Southern Lazio, DICEM},
    city={Cassino},
    country={Italy}
}
\affiliation[insubria]{
    organization={University of Insubria, Department of Science and Technology},
    city={Varese},
    country={Italy}
}
\affiliation[unitn]{
    organization={University of Trento, Department of Industrial Engineering},
    city={Trento},
    country={Italy}
}
\affiliation[infnpd]{
    organization={Istituto Nazionale di Fisica Nucleare Sezione di Padova},
    city={Padova},
    country={Italy}
}
\affiliation[infnmi]{
    organization={Istituto Nazionale di Fisica Nucleare, Sezione di Milano},
    city={Milano},
    country={Italy}
}
\affiliation[infnto]{
    organization={Istituto Nazionale di Fisica Nucleare, Sezione di Torino},
    city={Torino},
    country={Italy}
}
\affiliation[infnbo]{
    organization={Istituto Nazionale di Fisica Nucleare, Sezione di Bologna},
    city={Bologna},
    country={Italy}
}
\affiliation[tifpa]{
    organization={Istituto Nazionale di Fisica Nucleare TIFPA},
    city={Trento},
    country={Italy}
}
\affiliation[fnal]{
    organization={Fermi National Accelerator Laboratory},
    city={Batavia},
    country={Illinois, USA}
}
\affiliation[armenia]{
    organization={A. I. Alikhanyan National Science Laboratory (Yerevan Physics Institute)},
    city={Yerevan},
    country={Armenia}
}

\author[dfa,infnpd]{Michele Rignanese}
\author[cisas,infnpd]{Sabrina Ciarlantini}
\author[dfa,infnpd]{Caterina Pantouvakis\corref{cor1}}\ead{caterina.pantouvakis@phd.unipd.it}
\author[dfa,infnpd]{Alessandra Zingaretti}

\author[fnal]{Artur Apresyan}
\author[infnpd]{Patrizia Azzi}
\author[fnal,infnpd]{Nicola Bacchetta}
\author[dfa]{Chiara Bonini\fnref{present1}}
\author[dfa,infnpd]{Davide Chiappara\fnref{present2}}
\author[infnbo]{Davide Falchieri}
\author[infnto]{Sara Garbolino}
\author[armenia]{Aram Hayrapetyan}
\author[dfa,infnpd]{Serena Mattiazzo}
\author[unitn,tifpa]{Lucio Pancheri}
\author[dfa,infnpd]{Devis Pantano}
\author[infnto]{Angelo Rivetti}
\author[infnto]{Manuel Rolo}
\author[insubria,infnmi]{Romualdo Santoro}
\author[infnpd]{Rosario Turrisi}
\author[cassino,infnpd]{Jeffery Wyss}
\author[fnal]{Irene Zoi}
\author[dfa,infnpd]{Piero Giubilato}

\cortext[cor1]{Corresponding author}
\fntext[present1]{Present address: University of Padova CISAS, and INFN Padova, Italy}
\fntext[present2]{Present address: CERN, Genève, Switzerland}

\begin{abstract}
This work presents test beam characterization of the ARCADIA Main Demonstrator 3, a \SI{200}{\micro\meter} thick Fully Depleted MAPS developed using a custom LFoundry \SI{110}{\nano\meter} CIS process on a high-resistivity substrate. 
Measurements using a \SI{120}{\giga\eV} proton beam demonstrate a detection efficiency exceeding 99\% and a spatial resolution down to \SI{3.8}{\micro\meter}. The study evaluates cluster size, spatial resolution, and efficiency as a function of the threshold and reverse bias voltage.
\end{abstract}

\begin{keyword}
MAPS \sep Solid state detectors \sep Tracking 
\end{keyword}

\end{frontmatter}

\newpageafter{abstract}
\tableofcontents

% -----------------------------------------------------------------------------------
%                            Section 1 - Introduction
% -----------------------------------------------------------------------------------
\section{Introduction}
\label{sec:intro}
Monolithic Active Pixel Sensors (MAPS) are emerging as the future technology for vertex and tracking detectors in High Energy Physics (HEP), due to their low material budget and production costs compared to hybrid pixel ones. The largest implementation of MAPS is given by the current Inner Tracking System of the ALICE experiment \cite{ITS2_upgrade}, constituted by ALPIDE chips \cite{ALPIDE}. 
Recent technological progress has enabled a new generation of MAPS that meets the increasingly stringent requirements of current and next-generation HEP experiments \cite{snoeys2023monolithic}. For this reason, MAPS are now being foreseen as the technology for current detectors upgrade, as well as by upcoming experiments. Future implementations include the LHCb Mighty tracker (based on HV-CMOS MAPS technology) \cite{mighty_tracker}, the ePIC tracker \cite{eic_yellow_report}, and the ALICE ITS3 one \cite{its3_tdr}, and are considered the baseline technology for future experiments at colliders, such as the FCC-ee \cite{fcc_ee}.
A key development is given by Fully Depleted MAPS (FD-MAPS), where a high reverse bias voltage allows the full depletion of the entire substrate, leading to faster and more efficient charge collection, dominated by drift rather than diffusion.
This work focuses on the ARCADIA Main Demonstrator 3 (MD3) tracking performance, evaluated with a \SI{120}{\giga\electronvolt} proton beam using a three-planes setup made of ARCADIA MD3 chips.
In \autoref{sec:sensor}, a brief description of the ARCADIA MD3 chip, the latest FD-MAPS prototype developed by the ARCADIA INFN collaboration, is given. The test beam setup and the measurements performed are described in \autoref{sec:measurements}. \autoref{sec:data_analysis} briefly describes the framework used for the data analysis, and results are reported in \autoref{sec:results}.

% -----------------------------------------------------------------------------------
%                            Section 2 - ARCADIA brief overview
% -----------------------------------------------------------------------------------
\section{ARCADIA MD3 overview}
\label{sec:sensor}
ARCADIA (Advanced Readout CMOS Architecture with Depleted Integrated sensor Arrays) is an FD-MAPS developed within an INFN collaboration and is implemented with a custom LFoundry \SI{110}{\nano\meter} CIS process using a high-resistivity substrate ($\uprho \geq$ \SI{2}{\kilo\ohm\cdot\centi\meter})\cite{da2025arcadia,110nm-sensor}.
The first fully functional prototype developed by the ARCADIA collaboration is the MD3, a chip that features a $512\times512$ pixel matrix with a square pixel pitch of \SI{25}{\micro\meter}, corresponding to an active area of $1.28\times1.28$ \si{\square\centi\meter}. The pixel matrix is segmented into 16 32-column wide independent sections, with parallel readout and independent tuning of operating parameters. Below the pixel matrix, the periphery containing all the biasing, control, and readout logic is implemented.
The in-pixel front-end electronics features a pre-amplifier based on a source-follower transistor configuration with capacitive amplification. The pre-amplifier is followed by a feedback branch, which sets the amplifier baseline and shaping time, and a common-source amplifier acting as a discriminator. More details and a schematic of the analog front-end circuit are given in \cite{twepp_proceedings}.
The ARCADIA MD3 features a digital, sparsified readout architecture. The chip outputs 32-bit data words serially, and each word contains an 8-bit timestamp provided by the clocked periphery. The data words are extended with additional information by the FPGA. In particular, the 8-bit chip timestamp is extended with the 24-bit one given by the FPGA using the same reference clock.\\
Threshold uniformity, energy and noise calibration, and charge collection efficiency of the ARCADIA MD3 have been measured, and the results are shown in \cite{arcadia_characterization}.

\subsection{Depletion voltage measurements}
\label{subsec:iv_measurements}
\begin{figure}[htbp]
    \centering
    \begin{subfigure}{0.49\linewidth}
        \centering
        \includegraphics[width=\linewidth]{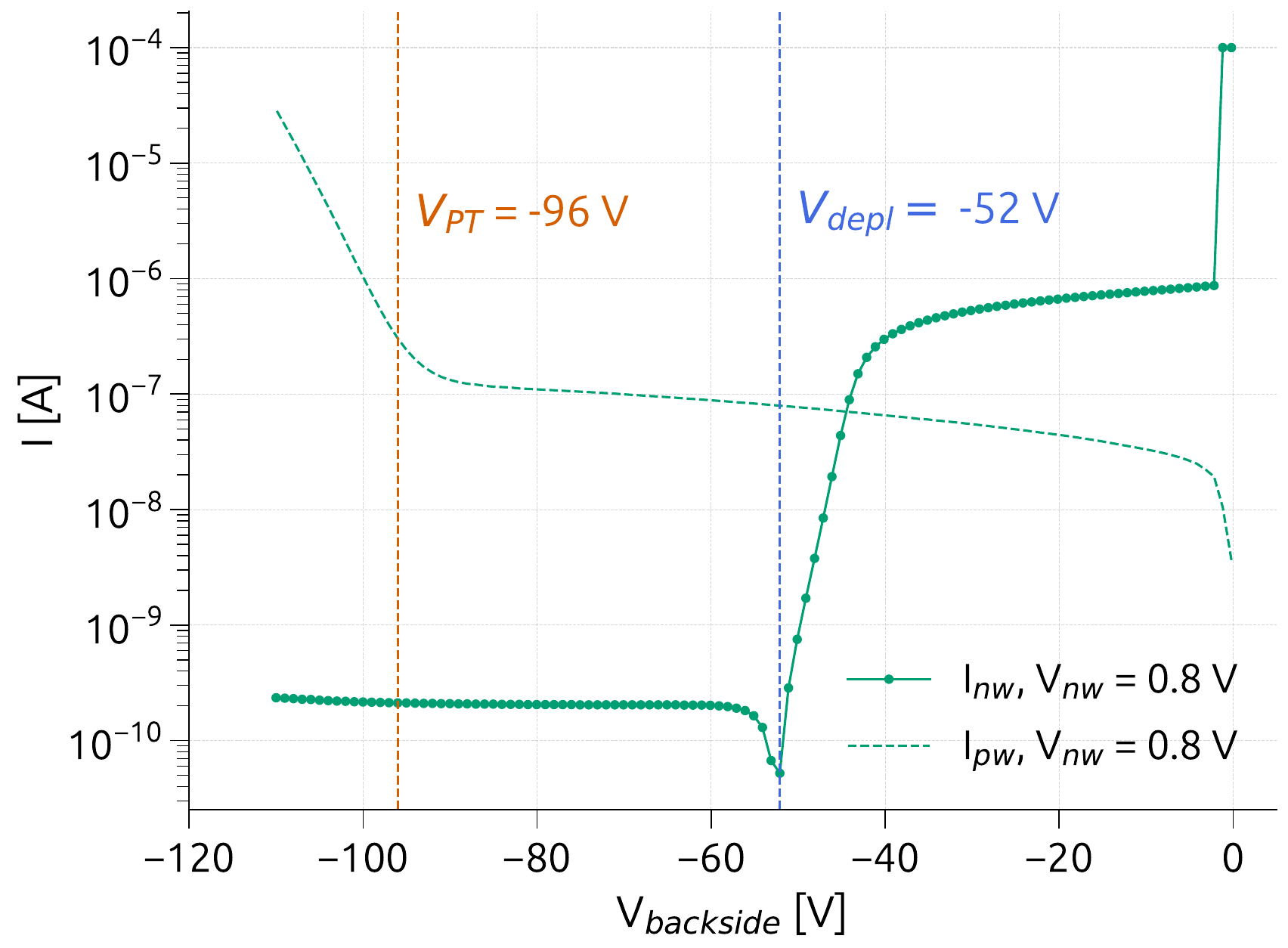}
	    \caption{}
	    \label{fig:pm25_iv}
    \end{subfigure}
    \hfill
    \begin{subfigure}{0.49\linewidth}
        \centering
        \includegraphics[width=\linewidth]{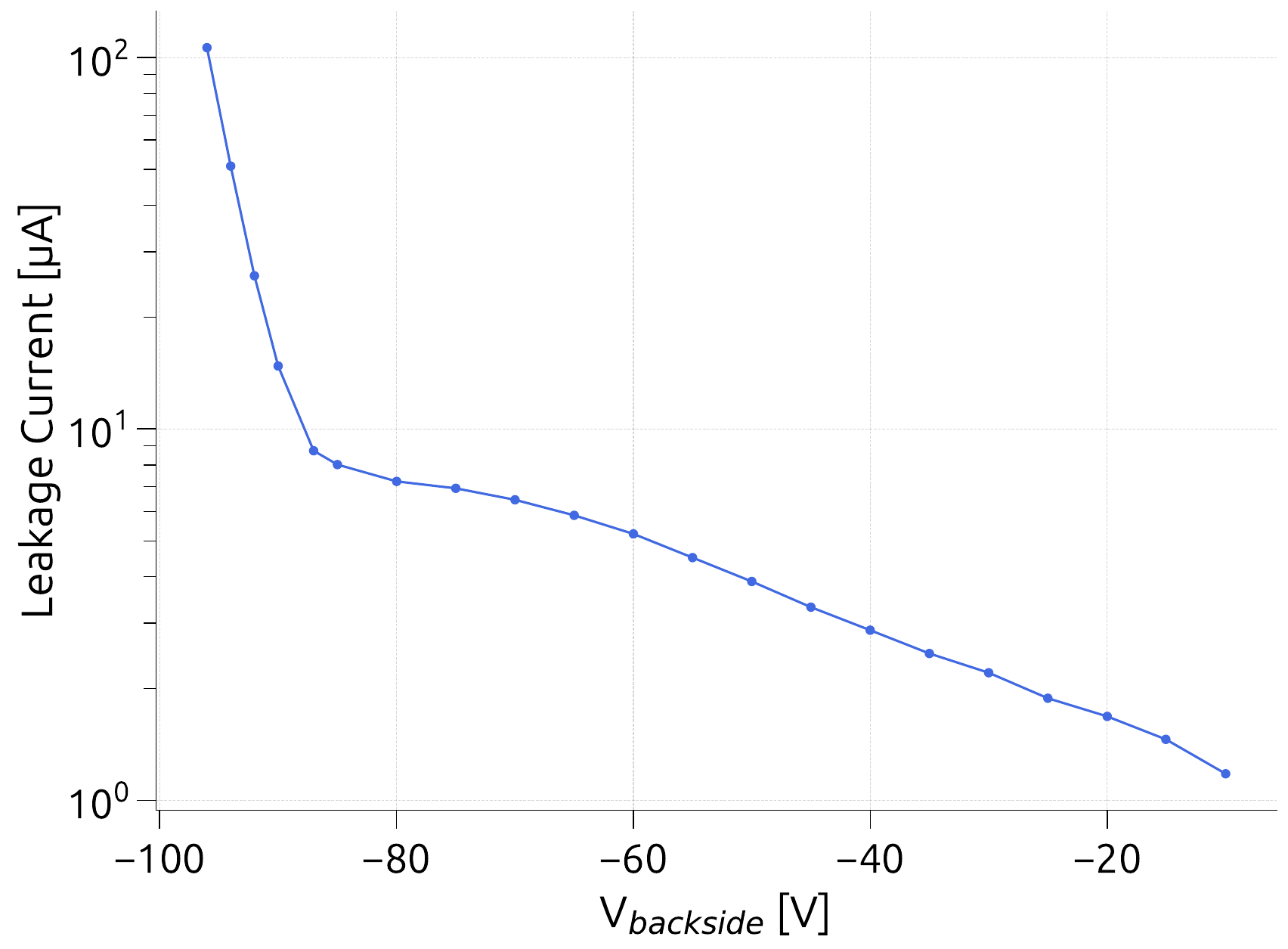}
	    \caption{}
	    \label{fig:md3_iv}
    \end{subfigure}
    \caption{\subref{fig:pm25_iv} IV characteristic of the passive test structure, a 60$\times$60 pixel array with the same MD3 electrode geometry and all n-well sensor electrodes connected in parallel, with n-wells voltage at \SI{0.8}{\volt}. The plot highlights the full depletion voltage (V$_{depl}$) and the punch-through voltage (V$_{PT}$). \subref{fig:md3_iv} Leakage current as a function of the applied reverse bias voltage for the ARCADIA MD3 chip.}
    \label{fig:iv_measurements}
\end{figure}

The MD3 prototype, object of this study, is thinned down to \SI{200}{\micro\meter}. Complete depletion of the low doped n-type active region is achieved by applying a high bias voltage ($\mathcal{O}(\SI{100}{\volt})$ for \SI{200}{\micro\meter} thickness) to the p$^+$ implant, formed through backside processing, with respect to the n-well collection electrodes, which are kept at \SI{800}{\milli\volt}. \\
To determine the optimal value ensuring the full depletion of the \SI{200}{\micro\meter}-thick substrate, the reverse bias voltage is evaluated based on IV measurements performed on both dedicated passive test structures \cite{110nm-sensor}, reported in \autoref{fig:pm25_iv}, and also on the MD3 itself (see \autoref{fig:md3_iv}).

Figure \ref{fig:pm25_iv} shows both the sensor n-well current (I$_{nw}$) and the deep p-well current (I$_{pw}$) as a function of the reverse bias voltage. The dip in the n-well current is assumed as the reference value for the depletion voltage (V$_{depl}$), while the steep increase in the p-well current represents the onset of the punch-through (V$_{PT}$) between the deep p-well and p$^+$ backside implant.
The same trend can also be observed in the leakage current of \autoref{fig:md3_iv}.
Considering the results from both measurements reported in \autoref{fig:iv_measurements}, the reverse bias voltage was set to \SI{-90}{\volt}. This value is well above the depletion voltage (\SI{-52}{\volt}) and below the punch-through voltage (\SI{-96}{\volt}), guaranteeing the full depletion of the substrate while maintaining the leakage current around \SI{10}{\micro\ampere} for the full MD3 chip.

% -----------------------------------------------------------------------------------
%                               Section 3 - Test beam measurements
% -----------------------------------------------------------------------------------
\section{Test beam measurements}
\label{sec:measurements}
In order to assess both the spatial resolution and the detection efficiency of the ARCADIA MD3 chip, several measurements were performed by varying the pixel threshold as well as the substrate depletion voltage.
The front-end parameter VCASN is the bias voltage of the input transistor of the discriminator,
which is controlled using a 6-bit DAC with a step of \SI{5}{\milli\volt}. This bias voltage is the main parameter that allows for the control of the pixel threshold, by setting the amplifier output baseline, and it is proportional to its complement.
For this reason, to ease data interpretation, from now on, the effective threshold will be expressed in DAC units as the complement of the VCASN parameter, i.e., as (63 - VCASN), so that the x-axis is directly proportional to the effective threshold.
In addition to the front-end bias voltage VCASN, there are other front-end bias currents which have an impact on the effective pixel threshold.
More details on how these three currents affect the pixel front-end response are given in \cite{twepp_proceedings}.

\subsection{Fermilab Test Beam Facility}
\label{subsec:ftbf}
The Fermilab Test Beam Facility (FTBF) \cite{ftbf-site} provides a spill-structured beam, with a spill duration of \SI{4.2}{\second} and a repetition rate of one spill per minute. 
The Main Injector (MI) cycle, continuously repeated in a spill, is \SI{11.2}{\micro\second} long and further divided into 7 batches, each \SI{1.6}{\micro\second} long.
Only the first batch is filled with protons. 
In the condition of smoothly extracted beam, only one particle per MI cycle is extracted per rotation. For intensities up to \SI{100}{\kilo\hertz}, double occupancy occurs 35\% of the time, with two particles per MI cycle extracted\footnote{\url{https://ftbf.fnal.gov/beam-delivery-path/}}.

\subsection{Experimental setup}
\label{subsec:telescope}
\begin{figure}[ht]
    \centering
    \begin{subfigure}{0.49\linewidth}
        \centering
        \includegraphics[height=5cm, trim={5cm 5cm 0 2cm},clip]{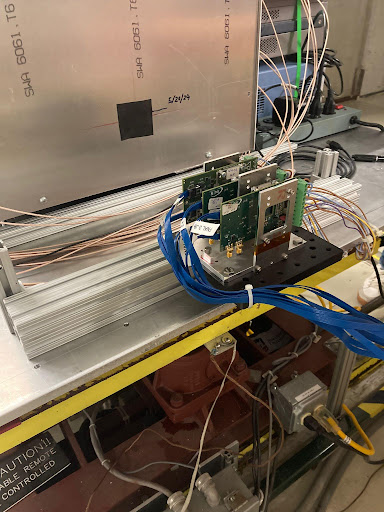}
	    \caption{}
	    \label{fig:telescope}
	\end{subfigure}
    \hfill
    \begin{subfigure}{0.49\linewidth}
        \centering
        \includegraphics[height=5cm, trim={3cm 1cm 3cm 1cm},clip]{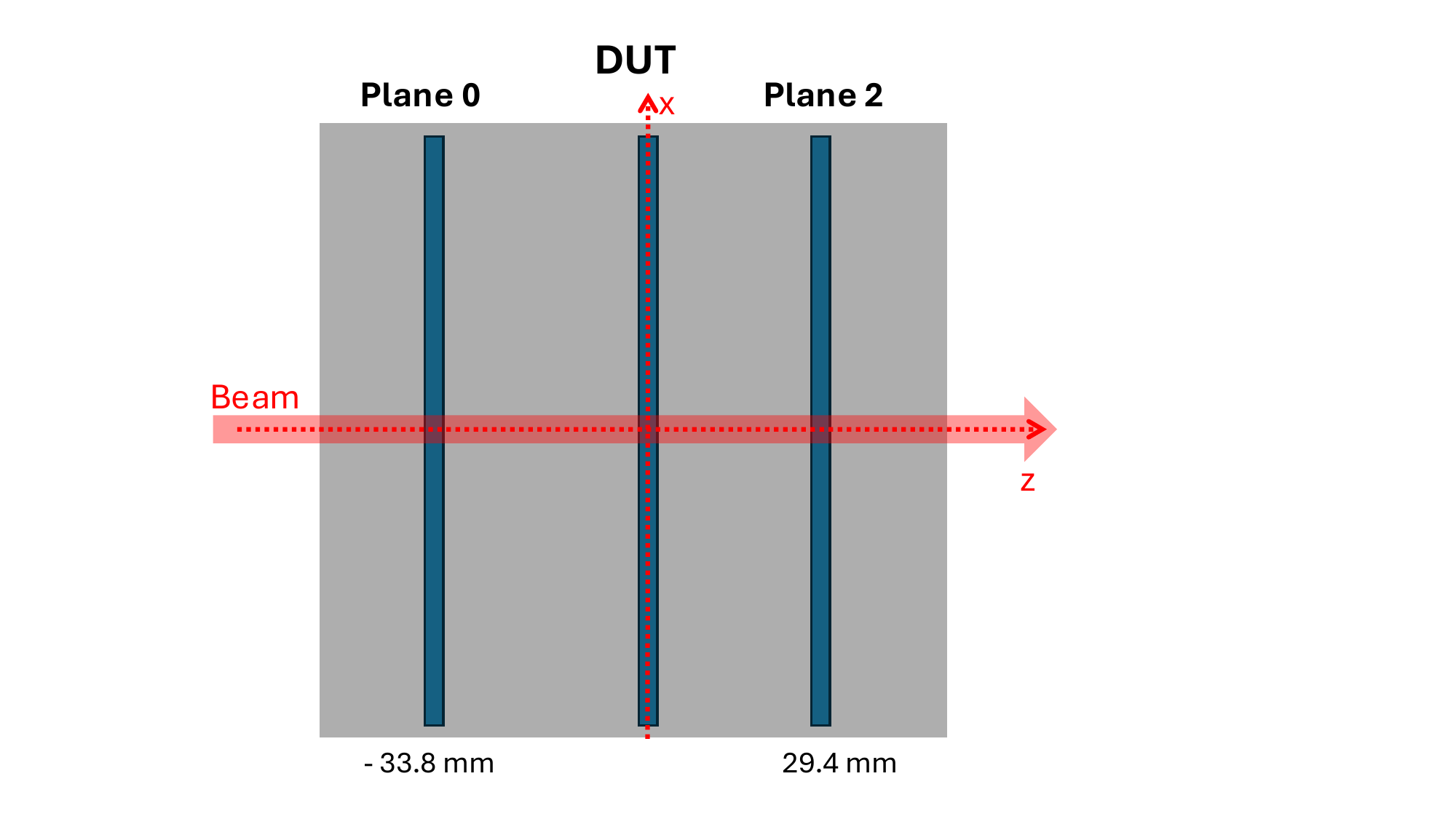}
	    \caption{}
	    \label{fig:telescope-scheme}
	\end{subfigure}
    \caption{\subref{fig:telescope} Picture of the telescope made of three ARCADIA MD3 planes used at the FTBF. \subref{fig:telescope-scheme} Telescope schematic: the dotted red lines represent the axes of the reference system, with the center corresponding to the center of the DUT.}
    \label{fig:telescope_picture_scheme}
\end{figure}

The setup is made of three \SI{200}{\micro\meter}-thick ARCADIA MD3 planes, i.e., one Device Under Test (DUT) and two external tracking planes, as shown in \autoref{fig:telescope}.

The DUT center is taken as the origin of the reference system, and the external tracking planes are placed along the beam axis: Plane 0 (P0) is at a distance of \SI{-33.8}{\milli\meter}, and Plane 2 (P2) at a distance of \SI{29.4}{\milli\meter}. The telescope scheme with the reference system and the distances between the planes is shown in \autoref{fig:telescope-scheme}. \\ 
In all measurements, a static mask of noisy pixels was applied. Specifically, 4, 1, and 12 noisy pixels of the first tracking plane, second tracking plane, and DUT, respectively, were masked during data taking, corresponding to less than 0.005\% of the total number of pixels.

\subsection{DAQ system and planes synchronization}
\label{subsec:synchronization}
\begin{figure}[ht]
    \centering
    \includegraphics[width=0.7\linewidth]{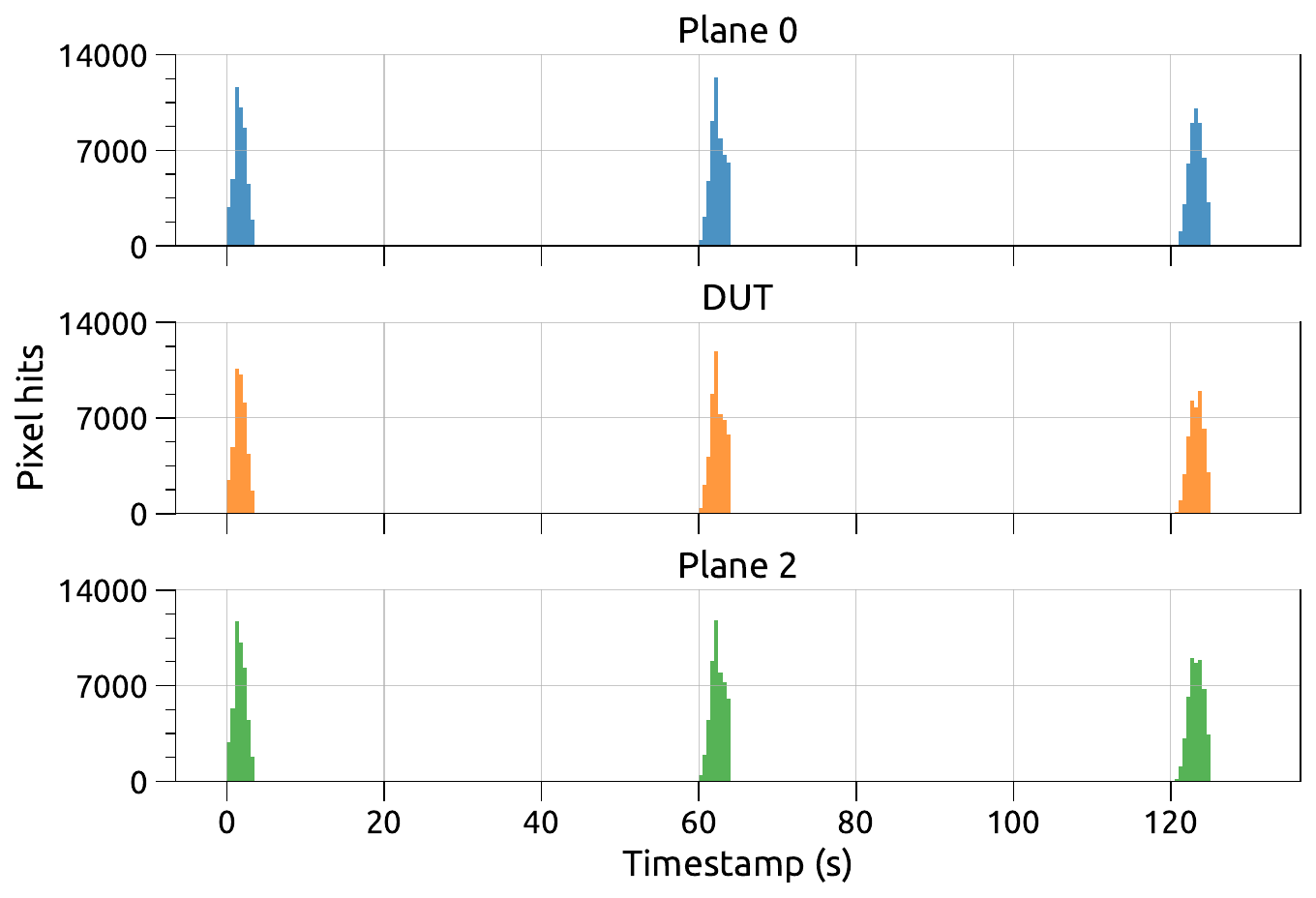}
    \caption{Pixel hit distributions over time (timestamp in seconds) for the three planes. The synchronization of the three devices is confirmed by the simultaneous arrival of the hits from all of them. These distributions reflect the FTBF beam structure: hits are recorded by all the planes every \SI{60}{\second} for a time window of approximately \SI{4}{\second}.} 
    \label{fig:data_structure}
\end{figure}

The Data AcQuisition (DAQ) system is based on a KC705 FPGA Board \cite{xilinx_kc705}. The chips are connected using the two FMC (FPGA Mezzanine Card) connectors, as the FPGA provides enough programmable logic to control up to three devices independently. The single-device firmware architecture is triplicated, providing each device its own 24-bit timestamp counter. To allow for chip synchronization, all three counters are driven by the same reference clock.
Since no trigger system is implemented in the experimental setup, the planes must be properly synchronized to reconstruct events in coincidence. The synchronization of the three planes is achieved by using the FPGA reset timestamp signal, which is enabled using a software command and allows resetting the FPGA timestamp counters of the three chips simultaneously. With this synchronization procedure, the timestamps of the chips are neglected, and all the information on the time of the event is given by the 24-bit FPGA timestamp counter, whose resolution is set to \SI{200}{\nano\second}.
\autoref{fig:data_structure} shows the synchronization of hits among the three planes obtained by exploiting the FPGA reset timestamp signal.

% -----------------------------------------------------------------------------------
%                               Section 4 - Data analysis
% -----------------------------------------------------------------------------------

\section{Data analysis}
\label{sec:data_analysis}
\begin{figure}[ht]
    \centering
    \includegraphics[width=0.7\linewidth]{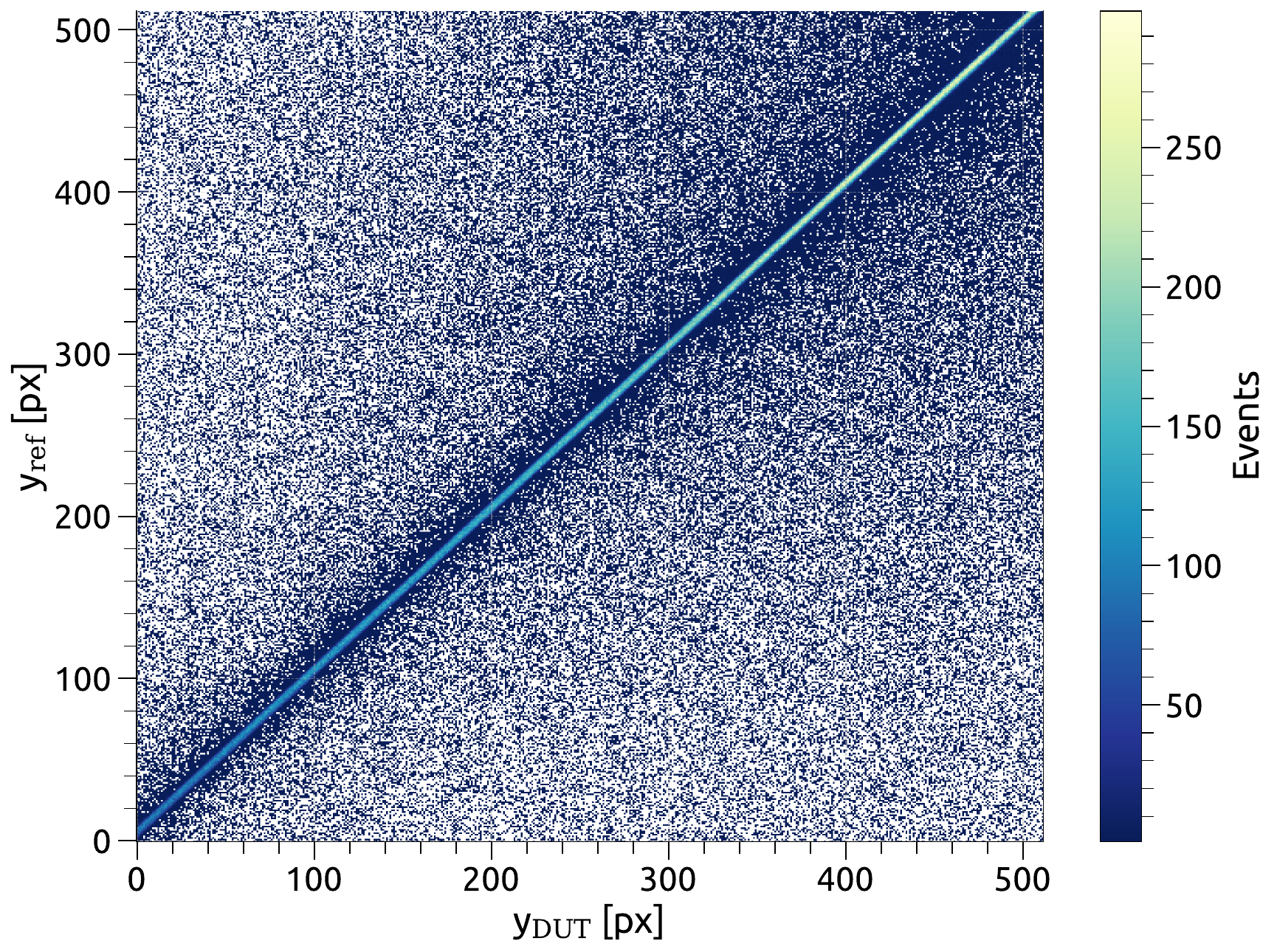}
    \caption{Correlation between the y coordinate of the DUT cluster and that of the reference-plane cluster.}
    \label{fig:correlation}
\end{figure}

After data decoding, all the analysis is done using the Corryvreckan framework \cite{corryvreckan}, a standard tool for test beam data analysis. After hits clusterization on each plane, the position of each cluster is reconstructed as the unweighted arithmetic mean of the centers of the pixels belonging to the cluster. Since the MD3 provides binary hit information, no pulse-height weighting is applied.
A first pre-alignment of the DUT and the P2 plane is performed using P0 as the fixed reference plane. The x- and y-shifts of the other two planes are extracted from the correlation histograms between each plane and the reference one. An example of a 2D plot of the correlations is shown in \autoref{fig:correlation}. 
With the position of the tracking planes constrained, a second more precise alignment procedure based on reconstructed tracks is performed to correct the DUT position for translational and rotational misalignment using residual distributions. 
Tracks on the external planes are reconstructed using the \textit{straight-line} model and associating pairs of clusters falling within a time window of \SI{5}{\micro\second}. Only events containing a single cluster on each tracking plane are retained.
Once the final geometric configuration is established, a spatial cut of 10 pixels is applied for the subsequent analysis to further constrain the identification of clusters on the two tracking planes corresponding to the same proton track by selecting only straight proton tracks, as it is unlikely that two clusters that do not correspond to the same particle track are within the set spatial window. This selection is necessary to avoid association ambiguities arising from multi-cluster events, since the setup includes only two tracking planes.
This leads to a fraction of about 79\% of the total clusters of the external planes that are used for track reconstruction.
The DUT clusters are then associated with the reconstructed tracks only if they fall within a time window of \SI{5}{\micro\second} and a spatial association radius of \SI{500}{\micro\meter}. An edge cut of 30 pixels is also applied in both rows and columns.

\section{Results}
\label{sec:results}
The following section presents the in-beam tracking performance of the ARCADIA MD3 chip. Three main features are analyzed and reported: cluster size, residual width, and detection efficiency. The results are presented as a function of the in-pixel threshold and as a function of the reverse bias voltage. \\
The cluster analysis is conducted considering only DUT clusters associated with tracks. The relevant quantities are the cluster size, i.e., cluster multiplicity, and the cluster width, which is defined as the maximum 1D elongation along the row or column axis. \\
As regards the residuals, the width of the distribution is taken as the $\upsigma$ parameter of the gaussian fit of the DUT residual histogram. \\
However, since the experimental setup consists of three identical sensors, the
tracking contribution can be mathematically computed and subtracted from the measured residual width, directly obtaining the true intrinsic ARCADIA MD3 spatial resolution (\autoref{subsec:intrinsic_res}).  
Consequently, all spatial performance results discussed in the following are corrected for the telescope contribution and are expressed directly in terms of the intrinsic spatial resolution of the device. \\
All the results reported in this section come with error bars, but given the very small magnitude of the uncertainties, they may not be visible in the plots.

\subsection{Resolution calculation}
\label{subsec:intrinsic_res}
Since no independent measurement of the spatial resolution of the ARCADIA MD3 is available, such quantity is extracted with a geometrical calculation.
In order to estimate the spatial resolution of the DUT an accurate determination of the tracking resolution is required. Since the setup consists of only three identical planes, there are two important consequences that must be noted.
First, the telescope’s contribution to the overall resolution is directly proportional to the intrinsic sensor resolution $\sigma_{det}$, which is the same for all three planes. Second, the tracks are reconstructed as straight lines connecting two points on the tracking planes, and then extrapolating the x-y position at the DUT z-position. On this basis, the spatial resolution of the DUT can be decoupled from the telescope contribution by following a simple geometrical calculation \cite{resolution_calculation}. \\
Considering the 1D case in the x-z plane represented in \autoref{fig:telescope-scheme}, the x position of the reconstructed hit on the DUT, x$_{DUT}$, can be computed using the equation of a straight line as a function of z for z = 0 (the DUT is at the center of the reference system). Given the measured positions x$_0$ on P0 (z$_0$ = \SI{-33.8}{\milli\meter}) and x$_2$ on P2 (z$_2$ = \SI{29.4}{\milli\meter}), x$_{DUT}$ can be computed as:

\begin{equation}
    x (z) = \dfrac{x_0z_2-x_2z_0}{z_2-z_0} + \dfrac{x_2-x_0}{z_2-z_0} \cdot z \underset{z_{DUT} = 0}{\longrightarrow} x_{DUT} = \frac{x_0 z_2 - x_2 z_0}{z_2-z_0} 
\end{equation}

Neglecting the uncertainties on z, the uncertainty on the extrapolation of x at the DUT position, i.e. the contribution of the telescope to the residuals along that direction, is then calculated as:

\begin{equation}
     \sigma_{tel, x}^2 = \frac{z_2^2+z_0^2}{(z_2-z_0)^2} \cdot \sigma_{det,x}^2
\end{equation}

The spatial resolution of the DUT can be therefore estimated in the following way:

\begin{equation}
    \sigma_{residual,x}^2 = \sigma_{det,x}^2 + \sigma_{tel,x}^2 = \sigma_{det,x}^2 + 0.502\,\sigma_{det,x}^2 \implies \sigma_{det,x} = \dfrac{\sigma_{residual,x}}{\sqrt{1.502}}
    \label{eq:intrinsic_res}
\end{equation}

The intrinsic sensor resolution $\sigma_{det}$ depends on the threshold, as can be seen in \autoref{fig:cluster_residual_width}. For the default threshold value of 1770 electrons (58 DAC), the mean spatial resolution is (3.775 $\pm$ 0.009) \si{\micro\meter}, where the uncertainty is only the statistical uncertainty given by the mean value of the gaussian fit.
The tracking planes P0 and P2 were operated at fixed threshold and reverse bias voltage settings chosen during the test beam to provide comparable detector response and leakage current to that of the DUT at its default setting of 1770 electrons. As a consistency check, the same analysis was repeated assigning to each tracking plane the role of DUT and tracking with the other two planes. The same geometrical treatment is applied, considering the specific positions of the planes depending on the configuration. The intrinsic spatial resolution of the first and second tracking planes is equal to 
(3.782 $\pm$ 0.009) \si{\micro\meter} and (3.797 $\pm$ 0.008) \si{\micro\meter}. The three values differ by less than 0.6\%, demonstrating a uniform behaviour of the three sensors.
Because the tracking planes were operated at the default configuration throughout all the measurements, their contribution to the pointing resolution of the telescope remains constant. Consequently, the geometrical correction of \autoref{eq:intrinsic_res}, can be systematically applied to the measured residual widths for all the subsequent results, allowing the performance to be expressed directly in terms of the intrinsic spatial resolution of the DUT.

\subsection{Threshold scan}
\label{subsec:thr_scan}
The behavior of cluster size, residual width, and detection efficiency is studied as a function of the threshold, which is varied in the range [38, 62] DAC. From the threshold calibration measurements of the ARCADIA MD3 \cite{arcadia_characterization}, the highest settable threshold, i.e., 62 DAC, corresponds to 2170 electrons, a value much smaller with respect to the signal of a Minimum Ionizing Particle (MIP) in \SI{200}{\micro\meter} of silicon ($\approx$ 16000 electrons).

\subsubsection{Cluster size and residual width}
\begin{figure*}[htbp]
    \centering
    \includegraphics[width=0.7\linewidth]{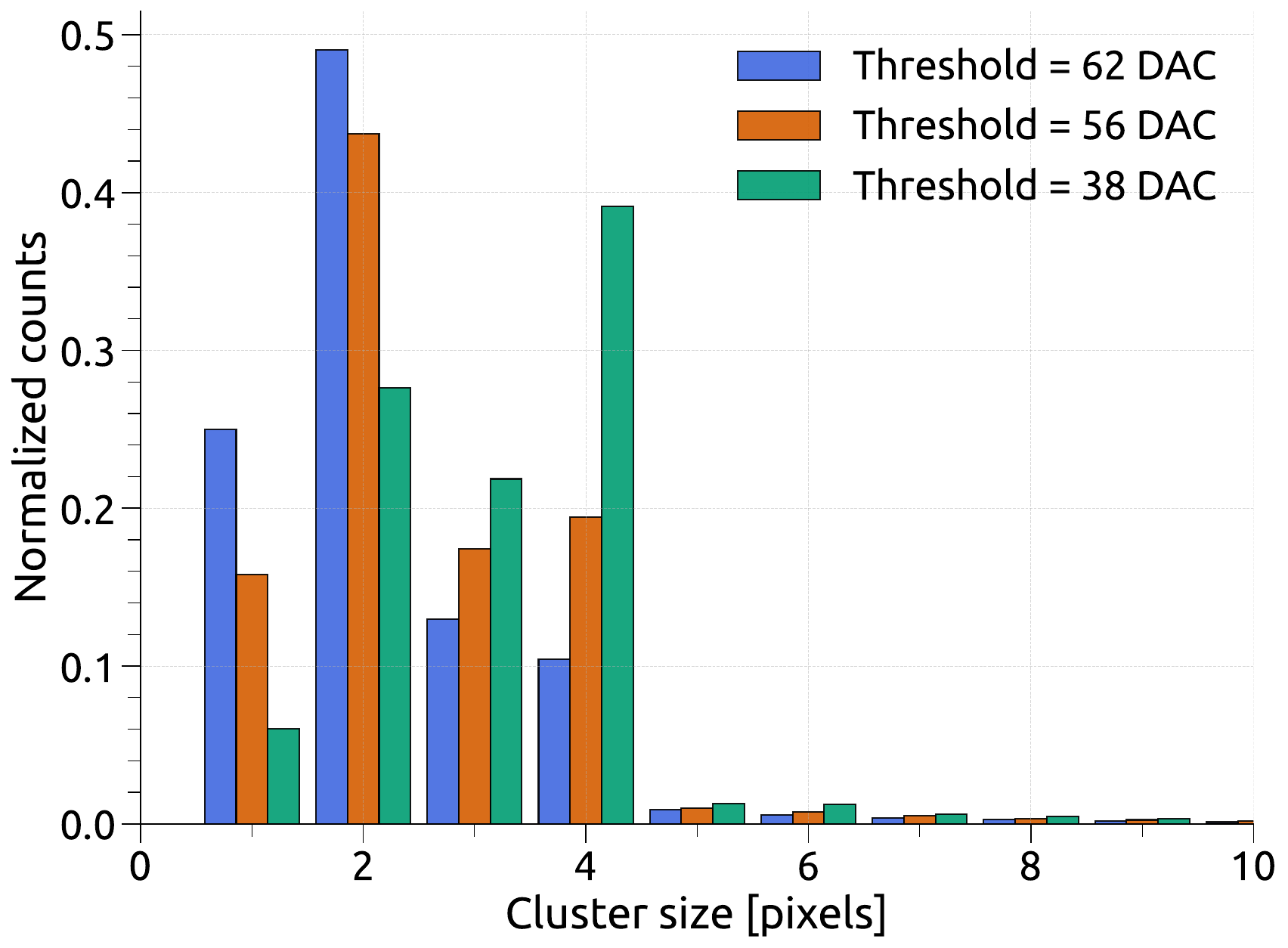}
    \caption{Comparison of the cluster size distribution at three different threshold values.}
    \label{fig:clz_size_comparison_threshold}  
\end{figure*}

The distribution of the cluster size for three different threshold values is shown in \autoref{fig:clz_size_comparison_threshold}. Since the sensor thickness is \SI{200}{\micro\meter}, charge sharing between adjacent pixels is expected to play a major role when collecting charge from the substrate.

\begin{figure*}
    \centering
    \includegraphics[width=0.75\linewidth]{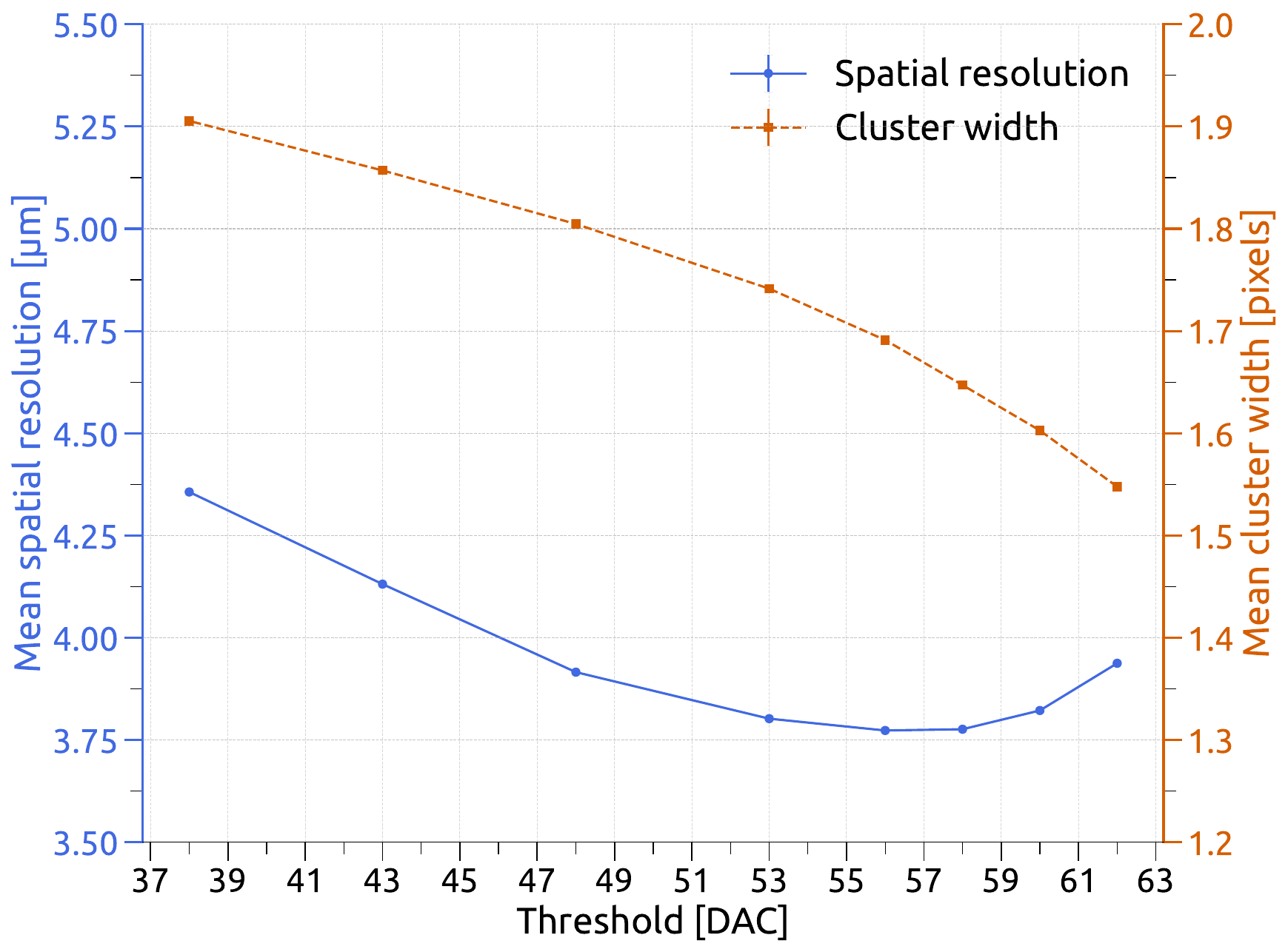}
    \caption{Average cluster width and spatial resolution as a function of the pixel threshold.}
    \label{fig:cluster_residual_width}
\end{figure*}

As a consequence, even at the highest threshold setting, clusters with a multiplicity larger than 1 are prevalent, indicating that the charge generated by a particle is often distributed between more pixels when the interaction occurs away from the pixel center. In fact, only 25\% of the clusters have a multiplicity equal to 1 at the highest threshold, while multiplicity 2 remains the dominant contribution. In the lowest threshold configuration, charge sharing is enhanced even more, and the most frequent multiplicity corresponds to 4 pixels. \\
In \autoref{fig:cluster_residual_width}, the average cluster width and the average width of the residual distribution between rows and columns are shown as a function of the threshold.
The cluster width increases as the threshold value decreases, and it is consistently greater than 1.5 pixels in the entire scanned threshold range.
For this reason, the residual width is always lower than the binary resolution of \SI{7.2}{\micro\meter}.

The minimum of the residuals is reached at a threshold value of 56 DAC, which corresponds to 1580 electrons, and has a mean width of \SI{4.65}{\micro\meter}.

\subsubsection{Fake hit rate and detection efficiency}
\begin{figure*}
    \centering
    \includegraphics[width=0.75\linewidth]{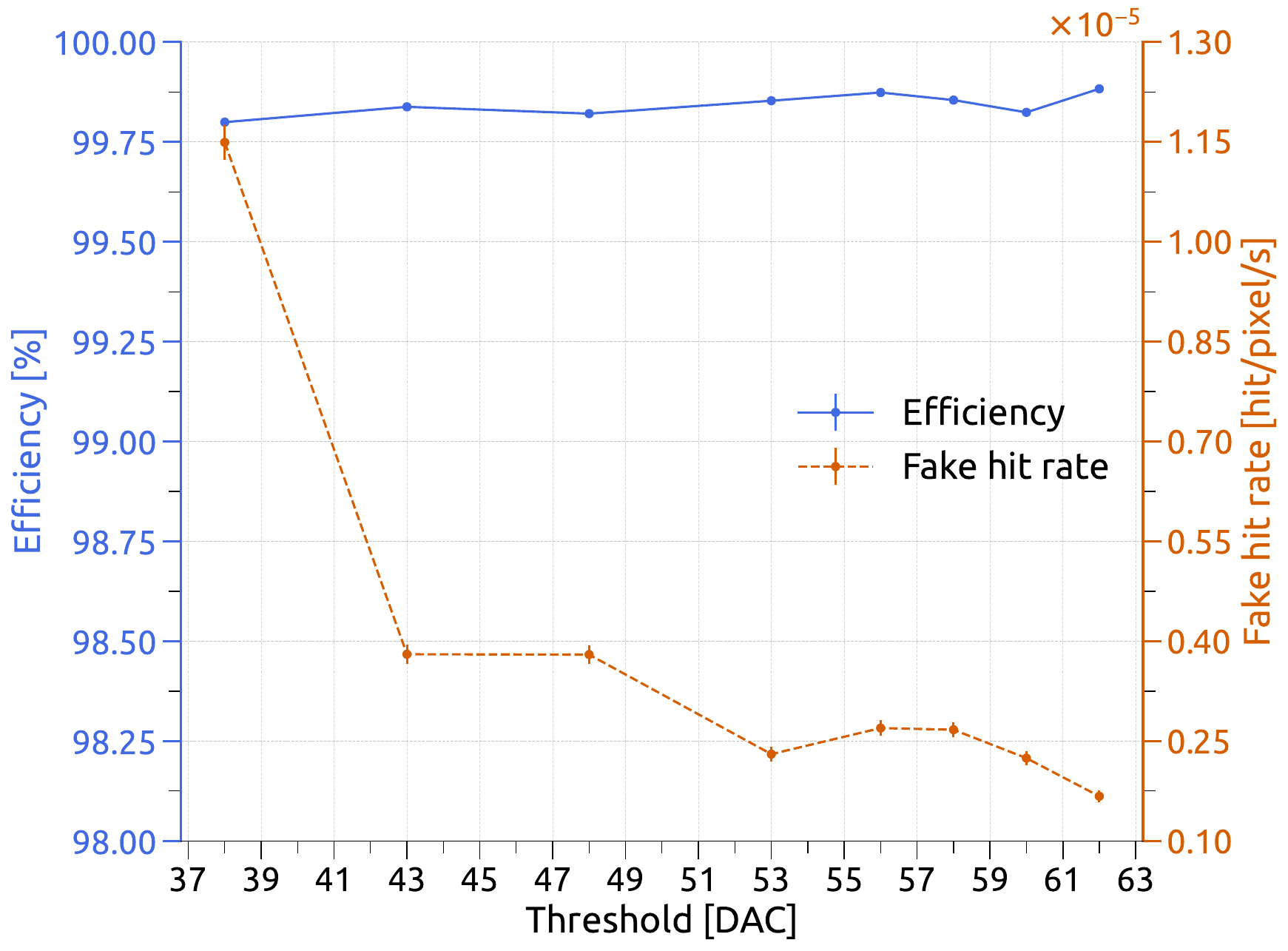}
    \caption{Detection efficiency and fake hit rate as a function of the pixel threshold.}
    \label{fig:eff_fhr}
\end{figure*}

The detection efficiency is greater than 99.8\% over the entire scanned threshold range, as shown in \autoref{fig:eff_fhr}, with a maximum variation of 0.2\%.
The weak dependence on threshold is consistent with the threshold remaining well below the signal expected from a MIP.
The fake hit rate, defined as the number of background hits per pixel per unit of time, was measured using the static mask for noisy pixels during no-beam time windows. It is lower than 4$\cdot$10$^{-6}$ hit/pixel/s for almost the entire threshold range. A major increase is visible only at the lowest threshold value of the scan.

\subsection{Reverse bias voltage scan}
\label{subsec:HV_scan}
\begin{figure*}[htbp]
    \begin{subfigure}{0.49\textwidth}
        \centering
        \includegraphics[width=\linewidth]{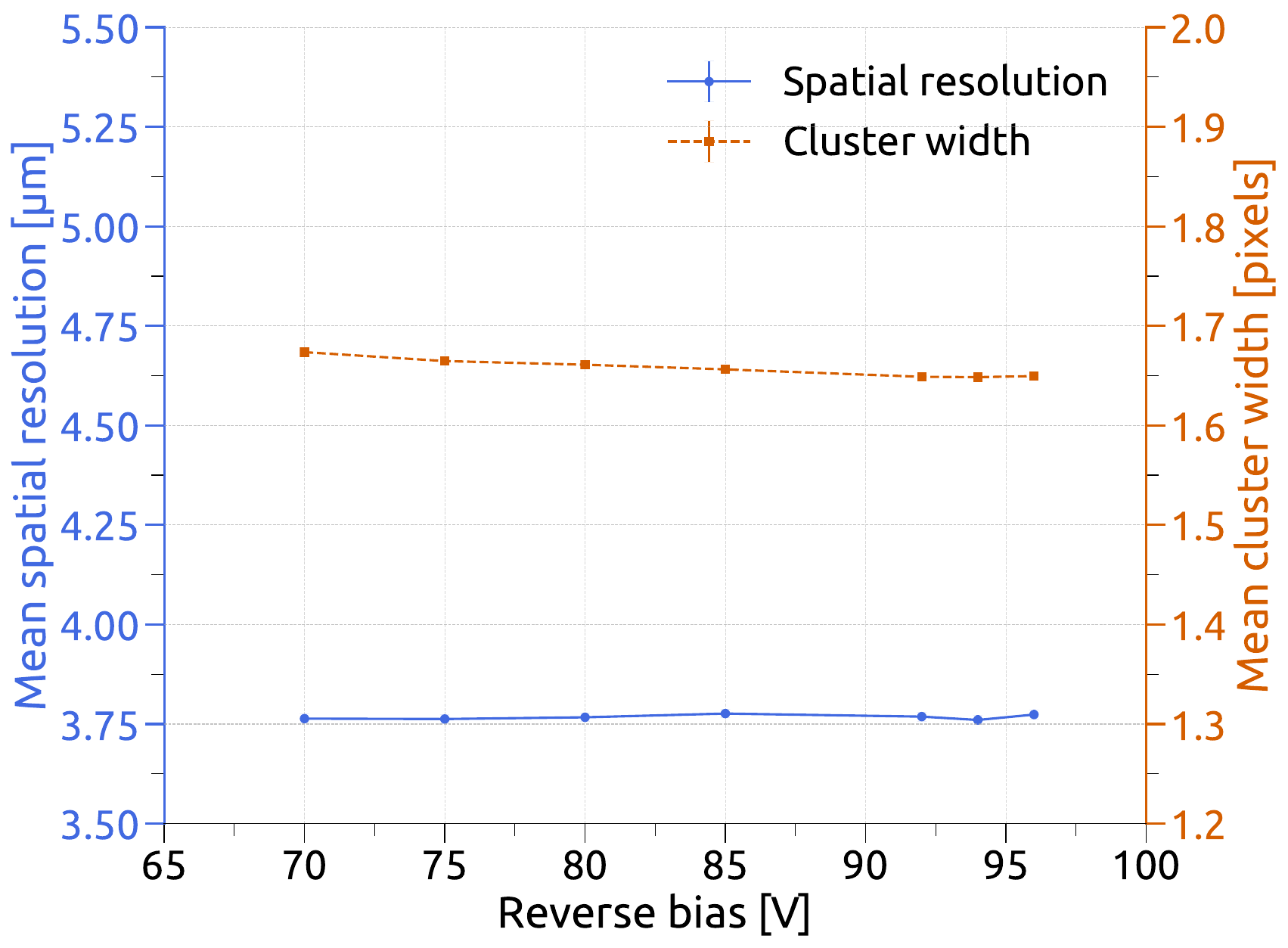}
        \caption{}
        \label{fig:res-hv}
    \end{subfigure}
    \hfill
    \begin{subfigure}{0.49\textwidth}
        \centering
        \includegraphics[width=\linewidth]{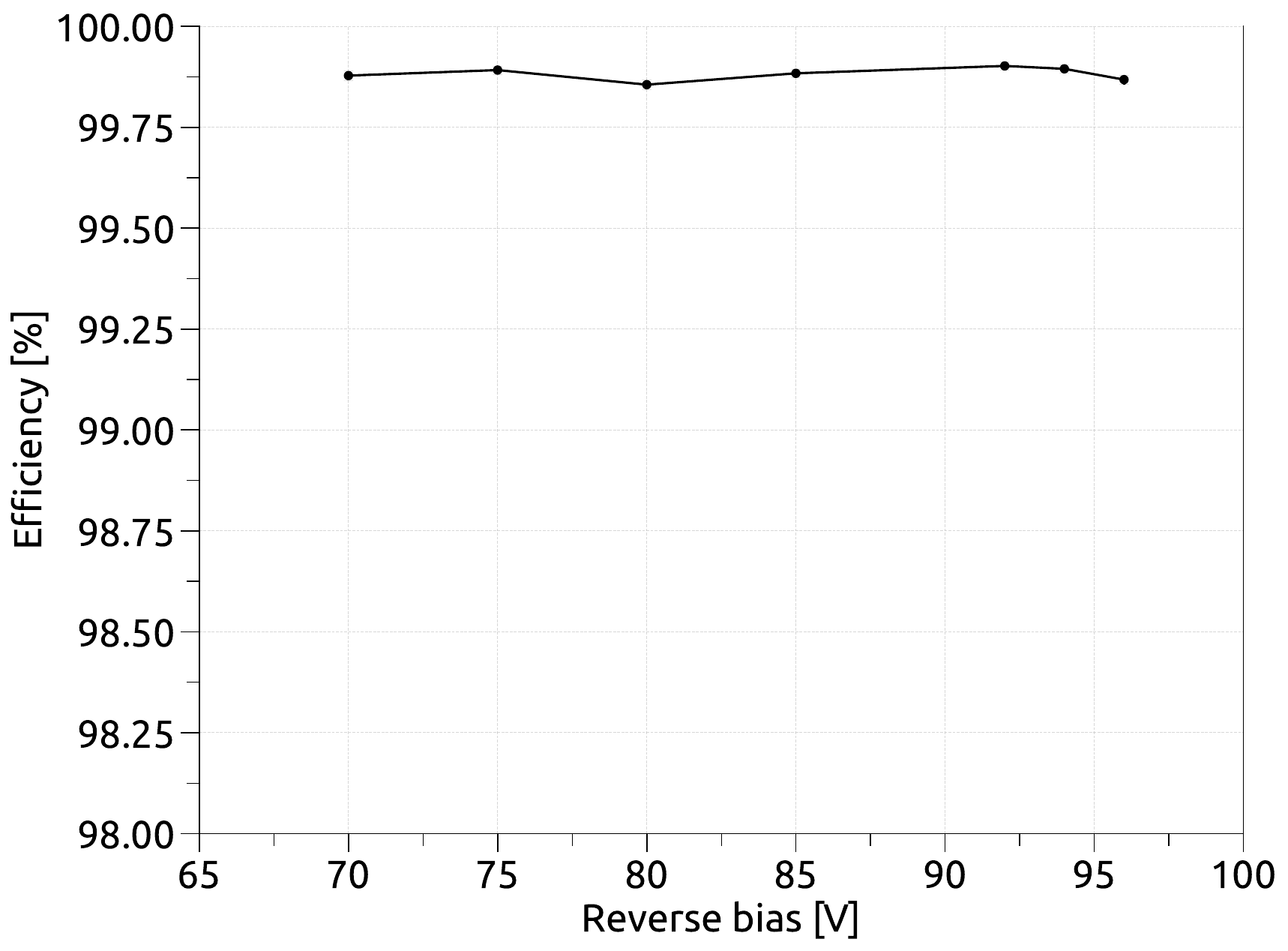}
        \caption{}
        \label{fig:eff-hv}
    \end{subfigure}
    \caption{Cluster and spatial resolution \subref{fig:res-hv} and efficiency \subref{fig:eff-hv} as a function of the sensor reverse bias voltage.}
    \label{fig:hv_scan_results}
\end{figure*}

The performance of the DUT is studied as a function of the reverse bias voltage in the range from \SIrange{-70}{-96}{\volt}, while keeping the DUT threshold at the default value of 1770 electrons (58 DAC). \\
Both the residual and cluster widths in \autoref{fig:res-hv} and the efficiency in \autoref{fig:eff-hv} show no significant dependence on the reverse bias voltage within the explored range. \\
As already mentioned in \autoref{sec:sensor}, the depletion of the substrate of the ARCADIA MD3 is achieved by applying a negative voltage between the backside p$^+$ implant and the top-side n-wells. The depletion region expands from the backside, and charge collection by drift is enabled only once the depletion region reaches the collection electrodes. Therefore, the signal at the electrode is generated only under the condition of full depletion. 
From IV measurements reported in \autoref{fig:iv_measurements}, it is possible to state that the full depletion of the substrate is reached at approximately \SI{-52}{\volt}. Even if the in-beam reverse bias voltage scan does not extend to lower reverse bias values to directly observe the transition from partial to full depletion, all the measured data (from \SIrange{-70}{-96}{\volt}) fall within the fully depleted range.
The results of this scan demonstrate that, once the substrate is fully depleted, there is no significant improvement or degradation in the tracking performance. \\
The maximum relative variation of the mean residual width is 0.7\%, while that of the detection efficiency is 0.2\%.

\subsection{Cluster size with tilted DUT}
\label{subsec:tilted_dut}
To evaluate the sensor response to non-orthogonal particle incidence, as was also done for the ALPIDE sensor for the ALICE Inner Tracking System upgrade \cite{bent_alpide}, the DUT was rotated by $7.5^{\circ}$ and $15^{\circ}$ around the axis corresponding to the pixel columns.
The inclined tracks increase the particles path length within the active substrate, resulting in a larger total deposited charge and a wider geometrical projection of the electron cloud.

\begin{figure}[ht]
    \centering
    \begin{subfigure}[]{0.49\textwidth}
        \centering
        \includegraphics[width=\linewidth]{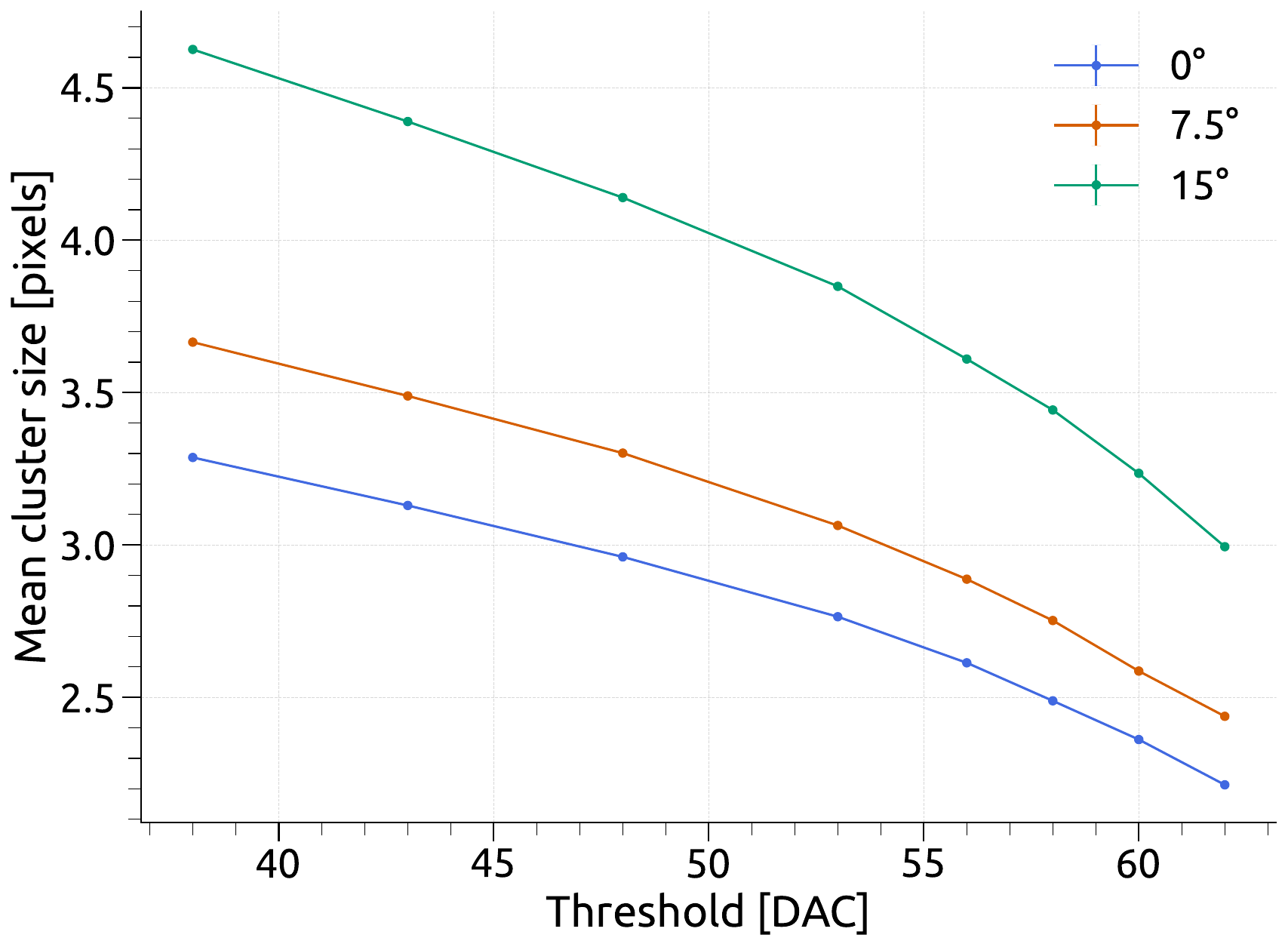}
        \caption{}
        \label{fig:fnal_clz_size_angles}
    \end{subfigure}
    \hfill
    \begin{subfigure}[]{0.49\textwidth}
        \centering
        \includegraphics[width=\linewidth]{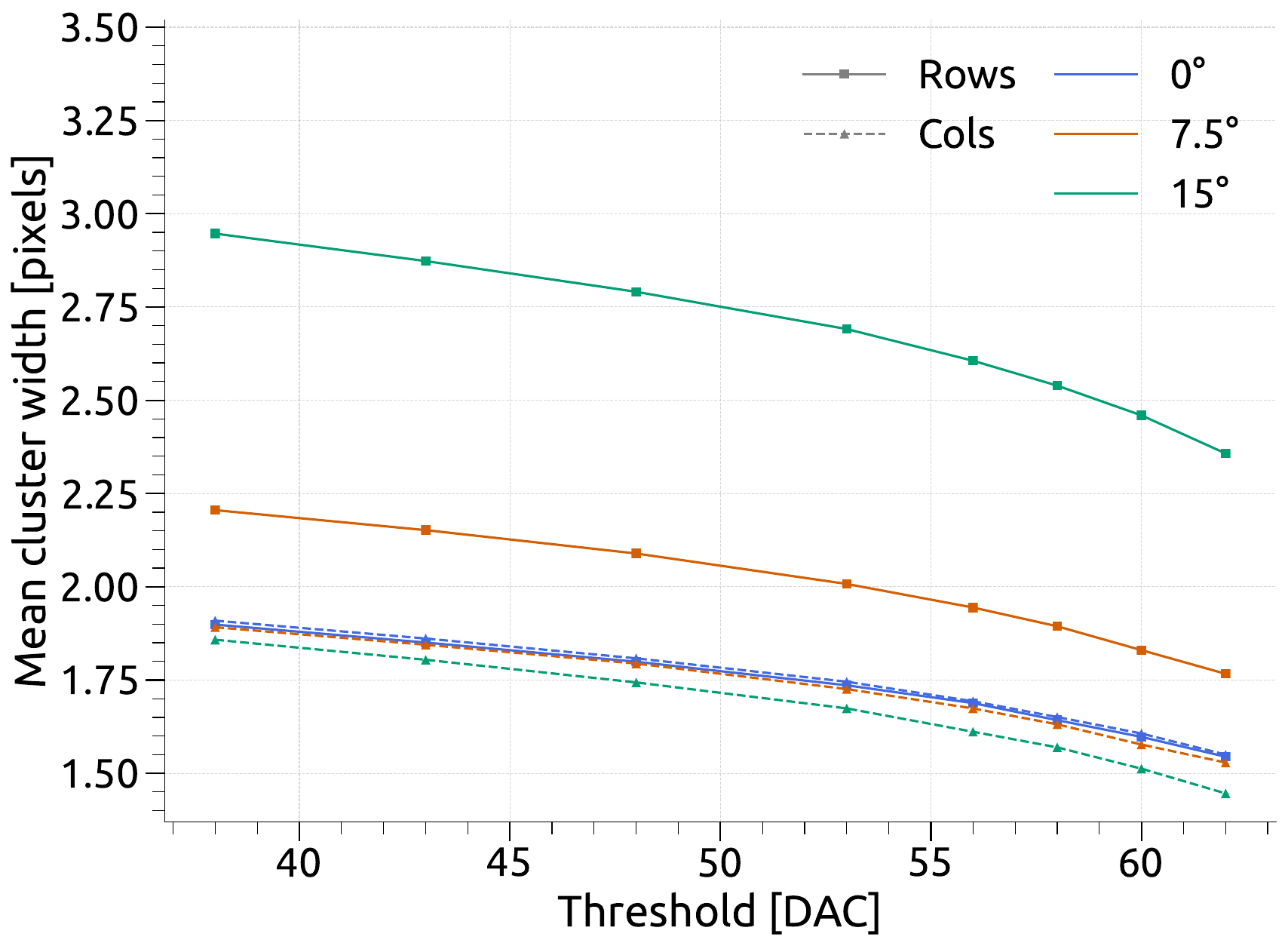}
        \caption{}
        \label{fig:fnal_clz_width_angles}
    \end{subfigure}
    \caption{(\subref{fig:fnal_clz_size_angles}) Mean cluster size as a function of the threshold for different DUT inclination angles. (\subref{fig:fnal_clz_width_angles}) Mean cluster width evaluated separately along the row and column axes. The geometrical elongation is visible exclusively along the rows, corresponding to the axis affected by the rotation.}
    \label{fig:fnal_angle_scan}
\end{figure}

As shown in \autoref{fig:fnal_clz_size_angles}, the mean cluster size systematically increases across the entire threshold scan range, growing by approximately 1.5 pixels at a $15^{\circ}$ tilt under nominal settings. As expected from the rotation geometry, a large variation of the cluster elongation is observed on the row axis (\autoref{fig:fnal_clz_width_angles}), while the mean cluster width along the columns remains almost identical to the $0^{\circ}$ baseline.

% -----------------------------------------------------------------------------------
%                                Section 6 - Conclusions
% -----------------------------------------------------------------------------------

\section{Conclusions}
\label{sec:conclusions}
The ARCADIA MD3 was tested, with a telescope made of two MD3 tracking planes, at the Fermilab Test Beam Facility with a \SI{120}{\giga\electronvolt} proton beam. \\
The average cluster width is always greater than 1.5 pixels due to charge sharing effect, explaining why the residual distribution width is lower than the binary resolution for all scanned parameters. The residual width is \SI{4.65}{\micro\meter} at its minimum. At the threshold value of 1770 electrons, the extracted intrinsic spatial resolution of ARCADIA MD3 is found to be \SI{3.8}{\micro\meter}.\\
The efficiency is found to be independent of the threshold and the reverse bias voltage, exceeding 99.8\% for both scanned parameters.

\section*{Acknowledgements}
The author(s) declare that financial support was received for the research and/or publication of this article. This work has received funding from INFN.\\ 
This study was carried out within the Space It Up project and received funding from the ASI and the MUR – Contract n. 2024-5-E.0 - CUP n. I53D24000060005.\\
This work was produced by FermiForward Discovery Group, LLC under Contract No. 89243024CSC000002 with the U.S. Department of Energy, Office of Science, Office of High Energy Physics. Publisher acknowledges the U.S. Government license to provide public access under the DOE Public Access Plan \href{https://www.energy.gov/downloads/doe-public-access-plan}{DOE Public Access Plan}.\\
This work was supported by the Science Committee of Republic of Armenia (Research projects No.22AA-1C009 and 22rl-037).

\section*{Declaration of generative AI and AI-assisted technologies in the manuscript preparation process}
During the preparation of this work the authors used Overleaf AI assistant tool for language editing and grammar check. After using this tool/service, the authors reviewed and edited the content as needed and take full responsibility for the content of the published article.

\bibliographystyle{elsarticle-num} 
\bibliography{biblio}

\end{document}